%% file: main.tex
\documentclass[amsmath,amssymb,aps,pre,twocolumn,groupedaddress, nobibnotes,floatfix]{revtex4-2}
\usepackage{graphicx}
\usepackage{color}
\usepackage{bm}
\usepackage{amsmath}
\usepackage{slashed}
\usepackage{mathrsfs}
\usepackage[normalem]{ulem}
\usepackage{comment}
\usepackage{booktabs}

\usepackage{dcolumn}
\usepackage{hyperref}
\hypersetup{colorlinks=true, citecolor=blue, urlcolor=blue, linkcolor=blue}
\usepackage{epstopdf}
\usepackage{subcaption}
\usepackage{caption}
\DeclareMathOperator{\sech}{sech}
\DeclareMathOperator{\arctanh}{arctanh}
\usepackage{extarrows}

\usepackage{tikz}
\usepackage{pgfplots}
\usepgfplotslibrary{colormaps} 
\pgfplotsset{compat=1.18}
\usetikzlibrary{calc}

\begin{document}

\title{Mean-Field Theory for Heider Balance under Heterogeneous Social Temperatures}

\author{Zhen Li}
\email{li-zhen@g.ecc.u-tokyo.ac.jp}
\affiliation{Department of Complexity Science and Engineering, Graduate School of Frontier Sciences, The University of Tokyo, Kashiwa 277-8561, Japan.}
\author{Yuki Izumida}
\email{Izumida@k.u-tokyo.ac.jp}
\affiliation{Department of Complexity Science and Engineering, Graduate School of Frontier Sciences, The University of Tokyo, Kashiwa 277-8561, Japan.}

\begin{abstract}
{  
    Heider balance theory provides a fundamental framework for understanding the formation of friendly and hostile relations in social networks.
    Existing stochastic formulations typically assume a uniform social temperature, implying that all interpersonal relations fluctuate with the same intensity.
    However, studies show that social interactions are highly heterogeneous, with broad variability in stability, volatility, and susceptibility to change.
    In this work, we introduce a generalized Heider balance model on a complete graph in which each link is assigned its own social temperature.
    Within a mean-field formulation, we derive a distribution-dependent self-consistency condition for the collective opinion state and identify the criteria governing the transition between polarized and non-polarized configurations.
    This framework reveals how the entire distribution of interaction heterogeneity shapes the macroscopic behavior of the system.
    We show that the functional form of the inverse temperature distribution, in particular whether it is light-tailed or heavy-tailed, leads to qualitatively distinct phase diagrams.
    We also establish universal bounds for the critical transition, where the homogeneous-temperature limit provides a universal lower bound for the critical mean of an inverse temperature distribution governing the transition.
    Numerical simulations confirm the theoretical predictions and highlight the nontrivial effects introduced by heterogeneity.
    Our results provide a unified route to understanding structural balance in realistic social systems and lay the groundwork for extensions incorporating fluctuations beyond mean field, external fields, and network topologies beyond the complete graph.
}
\end{abstract}
\maketitle

\section{Introduction}
Social relationships are shaped by both structural constraints and individual-level behavioral tendencies~\cite{Heider01011946,blum1985structural,cartwright1956structural,berk2000impact}.
A classical framework for formalizing such dynamics is Heider balance theory, which posits that triads of interpersonal relations tend to evolve toward a balanced configuration, minimizing cognitive and social tension~\cite{Heider01011946,cartwright1956structural}.
Heider balance comes from four simple and obvious principles~\cite{Heider01011946,cartwright1956structural,malarz2021heider}:
\begin{enumerate}
    \item Friends of my friends are my friends;
    \item Enemies of my friends are my enemies;
    \item Friends of my enemies are my enemies;
    \item Enemies of my enemies are my friends.
\end{enumerate}
These principles can be described by triadic balance on social networks~\cite{antal2005dynamics}.

Since the seminal works of Heider~\cite{Heider01011946}, and Cartwright and Harary~\cite{cartwright1956structural}, triadic balance has become a cornerstone concept in social groups~\cite{altafini2012dynamics,lan2020uncovering,corbetta2012intervention}.
The triadic balance may be considered on signed networks with plus (friend) or minus (enemy) sign being assigned to each link~\cite{zheng2015social}.
With the growing availability of large-scale networked data, modeling the collective evolution of such signed networks has gained renewed attention from statistical physics and sociophysics communities~\cite{malarz2021heider,malarz2022mean,rabbani2019mean,antal2005dynamics,bukina2025opinionformationisingsocial,burda2025relationshipheiderlinksising,burda2025heiderbalancesquarelattice,PhysRevLett.133.127402,belaza2017statistical,du2018reversing,belaza2019social,kirkley2019balance,bagherikalhor2021heider,babaee2022individual,hao2024proper,krawczyk2019heider,zhao2016bounded, stern2021impact}.

Beyond its sociological interpretation, Heider's balance mechanism is inherently a higher-order interaction process.
Such higher-order interaction is increasingly recognized as a central feature of real social and biological dynamics, where group interactions cannot be reduced to independent pairwise terms~\cite{lambiotte2019networks, battiston2021physics,schawe2022higher, xu2023linear, lambiotte2019networks, battiston2021physics, zhang2023higher}.

Recent advances have extended Heider's original deterministic formulation by introducing stochasticity and social temperature to reflect the variability of decision-making and the influence of external perturbations~\cite{malarz2022mean,rabbani2019mean,burda2025relationshipheiderlinksising,bagherikalhor2021heider,babaee2022individual,mohandas2025paradise,krawczyk2019heider}.
In particular, studies on temperature-driven Heider dynamics on complete graphs have demonstrated that changing noise level can induce discontinuous transitions between polarized and non-polarized states~\cite{malarz2022mean, burda2025relationshipheiderlinksising,bagherikalhor2021heider,babaee2022individual}.
However, these works typically assume a homogeneous social temperature, implying that all interpersonal relations fluctuate with the same intensity.

In real social networks, heterogeneity is ubiquitous~\cite{varshney2022social,dakin2020reciprocity}.
Studies show that interpersonal relations exhibit substantial heterogeneity~\cite{wu2025impact, johnson1983interdependence}.
Moreover, observed distributions are often broad or heavy-tailed~\cite{eom2015tail,enders2008long}.
Prior studies demonstrated that heterogeneity can fundamentally reshape dynamical transitions on networks~\cite{zhao2021structural, ferreri2014interplay,bagherikalhor2021heider,babaee2022individual}.

\begin{figure}
    \captionsetup[subfigure]{
        position=top
    }
    \begin{subfigure}{\columnwidth}
        \caption{}
        \input{fig}
        \label{fig:illustration_a}
    \end{subfigure}
    \begin{subfigure}{\columnwidth}
        \caption{}
        \input{figb}
        \label{fig:illustration_b}
    \end{subfigure}
    \caption{
        (a) Schematics of a complete graph for Heider balance.
        Each link is assigned with its own social temperature represented by different colors.
        (b) An example of the stochastic update rule for $x_{ij}$ in one triangle. 
    }
    \label{fig:illustration}
\end{figure}
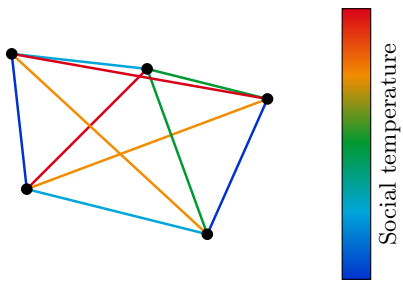

Motivated by these observations, we introduce in this work a generalized Heider balance model with \textit{heterogeneous social temperatures} assigned to each link of a network.
These temperatures determine the stochastic fluctuations of opinion variables through Boltzmann-like update probabilities.
The resulting dynamics can be viewed as a higher-order Ising-type system~\cite{alexandrov2025phasetransitionsisingmodel} with heterogeneous coupling strengths.
Building on this foundation, we establish a mean-field description that encapsulates how the distribution of inverse temperature governs macroscopic outcomes.
We highlight that the tail behavior of the inverse temperature distribution plays a decisive role in shaping the macroscopic phase structure.

The rest of the paper is organized as follows.
We introduce the Heider balance model with heterogeneous social temperatures in Sec.~\ref{sec_model}.
Then, we present the mean-field approach and some general properties in Sec.~\ref{sec_meanfield}.
Especially, the asymptotic behavior for the critical mean of the inverse temperature is provided there, which reveals the role of low temperature links and its ratio.
We also select different kinds of representative inverse temperature distributions as examples and discuss the corresponding phase diagrams in Sec.~\ref{sec_example}.
Finally, we address concluding remarks in Sec.~\ref{sec_con}.

\section{Heider balance model with heterogeneous social temperatures}
\label{sec_model}
Consider a social network consisting of $N$ nodes on a complete graph connected by undirected links (\autoref{fig:illustration_a}).
Each link between nodes $i$ and $j$ is characterized by an opinion variable $x_{ij}$, which can take values of $+1$ (positive) or $-1$ (negative).
The dynamics of the opinion variables $x_{ij}(t)$ towards Heider balance is governed by the following equation~\cite{antal2005dynamics, rabbani2019mean,malarz2021heider,malarz2022mean}:
\begin{equation}
    x_{ij}(t+1) = \mathrm{sgn}\left(\sum_{k \in \mathcal{M}_{ij}}x_{ik}(t)x_{kj}(t)\right),\label{eq:update_rule_deterministic}
\end{equation}
where $\mathcal{M}_{ij}$ is the set of common neighbors of nodes $i$ and $j$. It expresses the tendency for a pair of nodes to strengthen their mutual relationship if they have the same attitude toward their common neighbors.

To incorporate the effect of social temperature, we introduce a stochastic element to the update rule Eq.~\eqref{eq:update_rule_deterministic}, allowing for occasional deviations from the deterministic update, as illustrated by \autoref{fig:illustration_b}.
Specifically, we define the probability of updating the opinion variable $x_{ij}$ as follows:
\begin{equation}
    x_{ij}(t+1)=
    \begin{cases}
        +1, & \text{with probability } p_{ij} \\
        -1, & \text{with probability } 1- p_{ij}
    \end{cases},\label{eq:stochastic_rule}
\end{equation}
where 
\begin{equation}
    p_{ij} = \frac{\exp\left[T_{ij}^{-1}\xi_{ij}(t)\right]}{\exp\left[T_{ij}^{-1}\xi_{ij}(t)\right] + \exp\left[-T_{ij}^{-1}\xi_{ij}(t)\right]},\label{eq:stochastic_rule_p}
\end{equation}
and
\begin{equation}
    \xi_{ij}(t) = \sum_{k \in \mathcal{M}_{ij}} x_{ik}(t)x_{kj}(t).
\end{equation}
Here, $T_{ij}$ represents the social temperature associated with the link between nodes $i$ and $j$ (\autoref{fig:illustration_a}).
In this study, we consider heterogeneous social temperatures, where each link has its own temperature $T_{ij}$ drawn from a specified distribution.
A higher social temperature $T_{ij}$ indicates a greater likelihood of opinion changes, reflecting a more volatile social environment, while a lower social temperature suggests more stable opinions.
Specifically, in the limit $T_{ij} \to 0$, the update rule becomes deterministic as Eq.~\eqref{eq:update_rule_deterministic}, while in the limit $T_{ij} \to +\infty$, the opinion variable $x_{ij}$ is equally likely to be $+1$ or $-1$ regardless of the local configuration.

It is worth noting that the update rule defined above can be interpreted in the context of statistical mechanics.
The term $\xi_{ij}(t)$ can be interpreted as the local field acting on the opinion variable $x_{ij}$.
This field defines a corresponding local Hamiltonian $h(x_{ij})$ given by
\begin{align}
    h(x_{ij}) &= -x_{ij} \xi_{ij}= -x_{ij}\sum_{k \in \mathcal{M}_{ij}} x_{ik} x_{kj},\label{eq:local_hamiltonian}
\end{align}
where the sum runs over the set $\mathcal{M}_{ij}$ of nodes that form triads with the pair $(i,j)$.
The global Hamiltonian $\mathcal{H}$ for the whole system takes the form of~\cite{antal2005dynamics,rabbani2019mean,malarz2021heider,malarz2022mean}
\begin{equation}
    \mathcal{H} =\frac{1}{3}\sum_{(i,j)\in \mathcal{N}}h(x_{ij})=-\sum_{(i,j,k)\in \mathcal{T}} x_{ij} x_{jk} x_{ki},\label{eq:hamiltonian}
\end{equation}
which is similar to the Hamiltonian of an Ising model with higher-order interactions~\cite{robiglio2025higher}.
Here, the sum runs over all neighbors $(i,j)\in \mathcal{N}$ in the first equality, and all triads $(i,j,k)\in\mathcal{T}$ in the second equality, i.e., triples of nodes such that the three corresponding links $(i,j)$, $(j,k)$, and $(k,i)$ exist in the graph.

With this interpretation, the stochastic update rule of a single link [Eqs.~\eqref{eq:stochastic_rule} and \eqref{eq:stochastic_rule_p}] can be understood in analogy with Glauber dynamics~\cite{10.1063/1.1703954,antal2005dynamics}.
For a given local field $\xi_{ij}$, the two possible states $x_{ij}=\pm 1$ have the energy difference $2\xi_{ij}$, resulting in the Fermi-type probability $p_{ij}$.
In this sense, the Heider dynamics is formally analogous to the dynamics with a Glauber-type update rule for a spin subject to a local field generated by three-body interactions.

Unlike pairwise Ising-like models that encode only dyadic affinities, the fundamental update rule Eq.~\eqref{eq:update_rule_deterministic} or Eq.~\eqref{eq:stochastic_rule} explicitly depends on triadic motifs.
Each opinion variable $x_{ij}$ is influenced not by a single neighbor but by the collective configuration of all triangles involving nodes $i$ and $j$.
This makes the Hamiltonian Eq.~\eqref{eq:hamiltonian} a genuine three-body interaction system, structurally analogous to 
$p$--spin models with $p=3$~\cite{kirkpatrick1987p,robiglio2025higher}.

It is also worth noting that the present model does not possess a global spin-flip symmetry $x_{ij} \to -x_{ij}$.
The global spin flipping leads to $\mathcal{H} \to -\mathcal{H}$ for the global Hamiltonian Eq.~\eqref{eq:hamiltonian}.
This asymmetry implies that a state and the globally flipped state are not statistically equivalent and may differ in their interpretation and dynamical realization.
We can also find such difference in the self-consistent equation for the average opinion $\left<x\right>$ introduced in Sec.~\ref{sec_meanfield}.
As a consequence, the two branches corresponding to $\langle x \rangle > 0$ and $\langle x \rangle < 0$, where the latter does not exist as we can see in Sec.~\ref{sec_meanfield}, are not expected to play identical roles in the stationary state, and differs in their dynamical accessibility starting from random initial conditions.
In the mean-field formulation provided in Sec.~\ref{sec_meanfield}, only the branch $\langle x \rangle \ge 0$ appears as a solution.

\section{Mean-field approach and general properties}
\label{sec_meanfield}
\subsection{self-consistent equation}
\label{subsec_self}
The mean-field approach assumes that each opinion variable $x_{ij}$ experiences an average effect from the rest of the network.
By ignoring fluctuations and correlations between different opinion variables, we can derive a self-consistent equation for the average opinion $\left< x \right>$:
\begin{equation}
    \left< x \right> = \left< \tanh\left(\beta_{ij} \left< x \right>^2\right) \right>,\label{eq:self_consistent_0}
\end{equation}
where $\beta_{ij}\equiv (N-2)/T_{ij}$ captures the number of common neighborhood between node $i$ and $j$ in the graph and the social temperatures ($T_{ij}$) together.
Here, the average $\left<x\right>$ is taken over all the thermal average $\bar{x}_{ij}$ for a link between $i$ and $j$ in the network.
$\beta_{ij}$ works as heterogeneous ``coupling strengths'', which we may find in spin-glass systems~\cite{mezard1987spin,nishimori2001statistical}.
The effective coupling provides a compact representation of the force operating on each link.

We may consider that $\left<x\right>$ serves as an order parameter, which characterizes the collective state of the system.
A state with $\left< x \right> > 0$ refers to a polarized state, in which positive sign dominates.
In contrast, the state with $\left< x \right> = 0$ is referred to as a non-polarized state, where positive and negative relations are statistically balanced across the network.
We note that the self-consistent equation~\eqref{eq:self_consistent_0} admits only non-negative solutions for $\langle x \rangle$, which is consistent with the absence of global spin-flip symmetry.

Specifically, if we denote the distribution of the coupling strengths $\beta_{ij}$ as $\rho(\beta)$, the self-consistent equation~\eqref{eq:self_consistent_0} can be expressed as
\begin{equation}
    \left< x \right> = \int_0^{+\infty} \tanh\left(\beta \left< x \right>^2\right)\rho(\beta)d\beta,\label{eq:self_consistent}
\end{equation}
which holds one or three fixed points as can be easily shown.
We show the detailed derivation of the self-consistent equation~\eqref{eq:self_consistent} in Appendix~\ref{app:self}.

As consistent with previous study on the Heider balance model under complete graphs with homogeneous social temperatures~\cite{malarz2022mean}, the average opinion $\left<x\right>$ for models with heterogeneous social temperatures also undergoes a discontinuous phase transition from a polarized state $\left<x\right>\ne 0$ to a non-polarized state $\left<x\right>=0$ at a critical point determined by
\begin{equation}
    1 = \int_0^{+\infty} 2 \beta \left< x \right> \sech^2\left(\beta \left< x \right>^2\right)\rho(\beta)d\beta.\label{eq:critical_point}
\end{equation}

\subsection{Absence of hysteresis loop}
Similar as the case of the Heider balance model under a complete graph and homogeneous temperature~\cite{malarz2022mean}, there is no ``hysteresis loop'' when heating up and then cooling down the system by changing the mean of inverse social temperature $\mu$, regardless of the distribution.
When $\mu^{-1} \to 0^+$ implying that the system is fully cooled down, it is easy to find solutions $\left<x\right>=0$ and $\left<x\right>=1$, which lie on the trivial stable branch and non-trivial stable branch, respectively.
We may also assume a non-trivial unstable fixed point $\left<x\right>_{\rm u} = k\mu^{-1}$.
Substituting it into the self-consistent equation~\eqref{eq:self_consistent}, we have
\begin{align}
    k\mu^{-1} &= \int_0^{+\infty} \tanh\left(\beta k^2 \mu^{-2}\right)\rho(\beta)d\beta\notag\\
    &\simeq \int_0^{+\infty} \beta k^2 \mu^{-2}\rho(\beta)d\beta\notag\\
    &= k^2 \mu^{-1},
\end{align}
which leads to $k=0$ or $k=1$.
The case of $k=0$ corresponds to the trivial stable fixed point $\left<x\right>=0$, while the case of $k=1$ indicates the existence of a non-trivial unstable one when the system is fully cooled down.

Thus, the unstable fixed point $\left<x\right>_{\rm u}$ always exists even the mean temperature $\mu^{-1}$ is sufficiently low, indicating the absence of hysteresis loop when heating up and then cooling down the system.
It also consists with the previous study (Fig.~2 of~\cite{malarz2022mean}) on the Heider balance model under a complete graph with homogeneous social temperature, which can be regarded as a special case of our model with delta distribution of $\beta$.

\subsection{Universal bounds for the critical mean}
\label{subsec_bounds}
The critical mean $\mu_{\rm C}$ holds universal bounds applicable to any form of the distribution:
\begin{equation}
    C_* \leq \mu_{\rm C} < \frac{5}{4}\left(\frac{4}{3}\gamma^3\right)^{\frac{1}{5}},\label{eq:bounds}
\end{equation}
controlled by the third moment $\gamma^3$ of the distribution $\rho(\beta)$ in the valid range of $\gamma \geq \displaystyle\frac{25}{16}\sqrt{\frac{5}{3}}$.
Here, $C_*\approx 1.716$ is the critical value reached for a delta distribution~\cite{malarz2022mean}.
The derivation of the bounds~\eqref{eq:bounds} is provided in Appendix~\ref{app_bounds}

The lower bound indicates that the homogeneous temperature case achieves the highest critical temperature for the polarized state.
This can be understood in terms of the role of low temperature links in the heterogeneous setting, as discussed in Sec.~\ref{subsec_asym} and Sec.~\ref{sec_example}: a sufficiently large fraction of low temperature links enhances the stability of the polarized state.
In the homogeneous case, all links share the same (effectively low) temperature, corresponding to a fraction equal to one. As a result, the system can sustain the polarized state up to the highest critical temperature.

\subsection{Asymptotic behavior for the critical mean}
\label{subsec_asym}
Let $\lambda$ refers to the variance, higher-order moments, or other statistical properties of the distribution, the asymptotic behavior of the critical mean $\mu_{\rm C}$ as $\lambda \to +\infty$ depends on the behavior of the distribution:
\begin{eqnarray}
    \lim_{\lambda \to +\infty}\mu_{\rm C} =
    \begin{cases}
        +\infty &\quad \rho(\beta)~\text{vanishes}\\
        \text{finite} &\quad \rho(\beta)~\text{remains finite}
    \end{cases},\label{eq:asymptotic_mu}
\end{eqnarray}
The proof of Eq.~\eqref{eq:asymptotic_mu} is shown in Appendix~\ref{app_asym}

Importantly, this marks the essential difference between \textit{light-tailed} and \textit{heavy-tailed} distributions affecting the asymptotic behavior of critical $\mu$, when choosing $\lambda$ as variance or higher-order moments.
Light-tailed distributions vanish with diverging variance or higher-order moments, while heavy-tailed distributions could remain finite, resulting in different asymptotic behaviors of critical $\mu$ according to Eq.~\eqref{eq:asymptotic_mu}.
The qualitative difference between light-tailed and heavy-tailed distributions can be understood from the behavior of the inverse temperature distribution at large values.
While light-tailed distributions suppress the probability of low temperature links exponentially, heavy-tailed distributions maintain a finite fraction of such links even at large variance, thereby stabilizing the polarized phase with $\left<x\right>\ne 0$.

\section{Examples}
\label{sec_example}
To illustrate how heterogeneity reshapes the global phase diagram of the system, we examine several representative families of distributions $\rho(\beta)$.
These examples highlight how different forms of variability lead to qualitatively distinct phase diagram.
Light-tailed and heavy-tailed distributions each modify the location and structure of the transition between polarized and non-polarized states.
In particular, the behavior of higher-order moments, reflecting light-tailed or heavy-tailed property for distributions, can either destabilize polarized states or, conversely, maintain them even under strong noise level.

We here provide four examples with different distributions for $\beta$: \textit{delta distribution}, \textit{gamma distribution}, \textit{Pareto distribution}, and \textit{double-delta distribution}.
The delta distribution corresponds to fully homogeneous environments where all interpersonal relations fluctuate with identical stability.
The gamma distribution models light-tailed variability, typical of human activity patterns with finite variance and moderate deviations.
In contrast, the Pareto distribution represents heavy-tailed and scale-free heterogeneity, characteristic of systems where rare but extremely stable or influential ties dominate collective behavior.
Finally, the double-delta distribution illustrates how a sufficiently large fraction of low temperature links stabilizes the polarized state, thereby clarifying the key mechanism responsible for the qualitative differences in the phase diagrams of different temperature distributions.

\subsection{Delta distribution}
The delta distribution can represent the case with homogeneous social temperatures, i.e., $T_{ij}=T$ for all links.
In this case, the distribution of $\beta$ is given by
\begin{equation}
    \rho(\beta) = \delta\left(\beta - \mu\right),
\end{equation}
where $\mu=(N-2)/T$.
The self-consistent equation~\eqref{eq:self_consistent} and the critical point~\eqref{eq:critical_point} reduce to
\begin{align}
    \left< x \right> &= \tanh\left(\mu \left< x \right>^2\right),\\
    2 \mu \left< x \right> &=  \cosh^2\left(\mu \left< x \right>^2\right),
\end{align}
respectively, which consist with the previous result in~\cite{malarz2022mean}.
By solving these equations, we find that this homogeneous system shows a discontinuous phase transition at $\mu_{\rm C}\approx 1.716$~\cite{malarz2022mean}, which serves as the lower bound $C^*$ for any distribution (Eq.~\eqref{eq:bounds}).
    \begin{figure}
        \includegraphics[width=0.95\columnwidth]{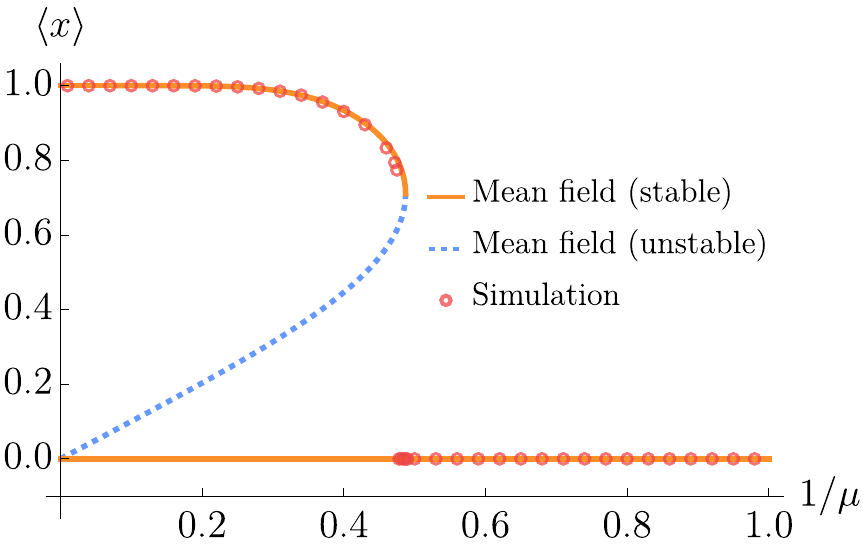}
        \centering
        \caption{
            Mean-field solutions of $\left<x\right>$ vs. $\mu^{-1}$ for $\sigma=1$ under the gamma distribution \eqref{eq:gamma}.
            The solid curve and line represent the stable solutions, while the dotted line represents the unstable solution. The critical point is approximately given by $\mu_{\rm C}^{-1}\approx 0.478$.
            The plots of $\mu^{-1}\in [0, 0.075]$ are linearly continued according to analytical properties shown in Sec.~\ref{sec_meanfield}.
            The red circles show the final 1000-step averages of 10000-step simulations of~\eqref{eq:stochastic_rule} on a complete graph with $N=100$, which agree with the mean-field results, giving the critical point as $\mu_{\rm C}^{-1}\approx 0.475$.
        }
        \label{fig:gamma_sigma}
    \end{figure}
    \begin{figure}
        \includegraphics[width=0.95\columnwidth]{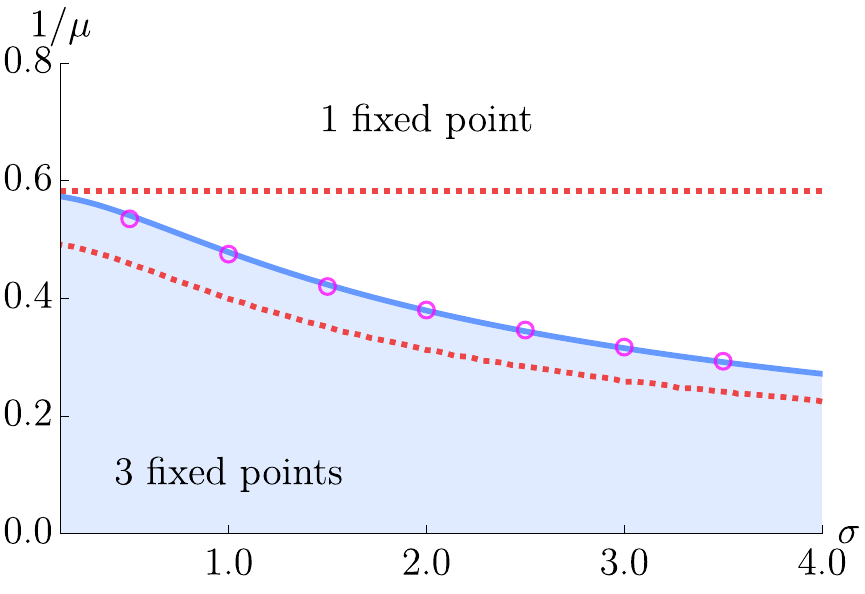}
        \centering
        \caption{
            Phase diagram with respect to $(\sigma,\mu^{-1})$ under the gamma distribution~\eqref{eq:gamma}.
            The diagram is separated by three fixed-points area and only one trivial fixed-point ($\left<x\right>=0$) area.
            The red-dotted curve and line show the upper and lower bounds for $\mu_{\rm C}$ in Eq.~\eqref{eq:bounds}, respectively, between which the critical curve separating the two areas is located.
            The magenta circles represent the simulation results of $\mu_{\rm C}$ on a complete graph with $N=100$.
        }
        \label{fig:gamma_phase}
    \end{figure}

\subsection{Gamma distribution}
The gamma distribution can represent the case with continuous and light-tailed inverse social temperatures.
In this case, the distribution of $\beta$ obeys:
\begin{equation}
    \rho(\beta) = \frac{\beta^{\alpha-1} e^{-\beta/\theta}}{\theta^{\alpha}\Gamma(\alpha)},\label{eq:gamma}
\end{equation}
where $\alpha>0$ and $\theta>0$ are the shape and scale parameters, respectively.
The gamma distribution has mean $\mu = \alpha \theta$ and standard deviation $\sigma = \sqrt{\alpha} \theta$.
Thus, we can express the shape and scale parameters as $\alpha = \mu^2/\sigma^2$ and $\theta = \sigma^2/\mu$, respectively.

\autoref{fig:gamma_sigma} shows the mean-field solutions of Eq.~\eqref{eq:self_consistent} as the function of $\mu^{-1}$ for $\sigma=1$ under the gamma distribution~\eqref{eq:gamma}. 
The system has a pair of stable and unstable polarized solutions $\left<x\right>\ne 0$ below a critical point, while the trivial non-polarized solution $\left<x\right>=0$ always exists. In accordance with the mean-field prediction for this heterogeneous system, the simulation results of~\eqref{eq:stochastic_rule} show the discontinuous phase transition from the polarized state to the non-polarized state at the critical point.

\autoref{fig:gamma_phase} shows the phase diagram with respect to $(\sigma, \mu^{-1})$ under the gamma distribution~\eqref{eq:gamma}.
The critical curve separating the three fixed-points area and the only one fixed-point area is located between the upper and lower bounds in Eq.~\eqref{eq:bounds}.
The upper bound line,
\begin{equation}
    \sigma^2 = \frac{\mu^2}{100}\left(\sqrt{\frac{3125+6144\mu^2}{5}}-75\right),\label{sigma_critical}
\end{equation}
is calculated from the right-hand side of Eq.~\eqref{eq:bounds} by considering $\gamma^3 = \alpha(\alpha+1)(\alpha+2)\theta^3$ for the gamma distribution.
When $\sigma \to +\infty$, the critical value approaches $0$, making the polarized phase vanish; the low temperature links obeying the light-tailed distribution are not sufficient to stabilize the polarized state against increasing $\sigma$.

\subsection{Pareto distribution}
The Pareto distribution can represent the case with scale-free inverse social temperatures in social networks.
In this case, the distribution of $\beta$ is given by
\begin{equation}
    \rho(\beta) = \frac{\alpha \beta_{\rm min}^{\alpha}}{\beta^{\alpha+1}},\label{eq:pareto}
\end{equation}
for $\beta \geq \beta_{\rm min}>0$, where $\alpha>0$ is the shape parameter.
The Pareto distribution has the mean $\mu = \dfrac{\alpha\beta_{\rm min}}{\alpha -1}$ and the standard deviation $\sigma = \dfrac{\beta_{\rm min}}{\alpha-1}\sqrt{\dfrac{\alpha}{\alpha-2}}$.

\begin{figure}
    \includegraphics[width=0.95\columnwidth]{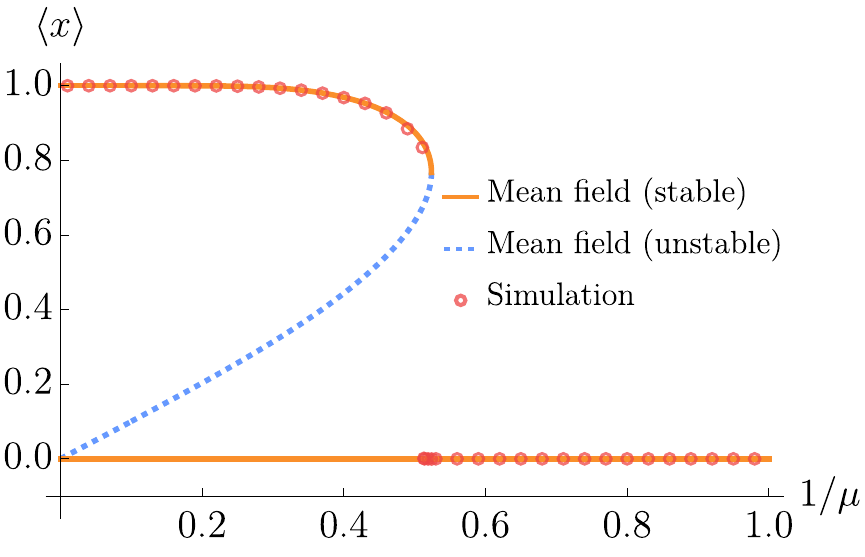}
    \centering
    \caption{
        Mean-field solutions of $\left<x\right>$ vs. $\mu^{-1}$ for $\sigma=1$ under the Pareto distribution~\eqref{eq:pareto}. The solid curve and line represent the stable solutions, while the dotted line represents the unstable solution. The critical point is approximately given by $\mu_{\rm C}^{-1}\approx 0.524$.
        The plots of $\mu^{-1}\in [0, 0.01]$ are linearly continued according to analytical properties shown in Sec.~\ref{sec_meanfield}.
        The red circles show the final 1000-step averages of 10000-step simulations on a complete graph with $N=100$, which agree with the mean-field results, giving the critical point as $\mu_{\rm C}^{-1}\approx 0.514$.}
    \label{fig:pareto_sigma}
\end{figure}
\begin{figure}
    \includegraphics[width=0.95\columnwidth]{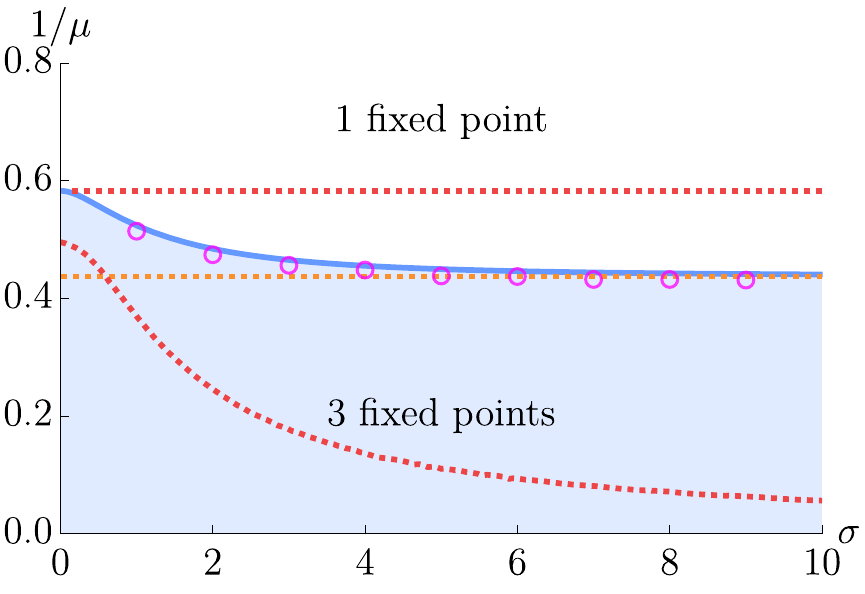}
    \centering
    \caption{
        Phase diagram with respect to $(\sigma, \mu^{-1})$ under the Pareto distribution~\eqref{eq:pareto}.
        The diagram is separated by three fixed-points area and only one trivial fixed-point ($\left<x\right>=0$) area.
        The red-dotted curve and line show the upper and lower bounds for $\mu_{\rm C}$ in Eq.~\eqref{eq:bounds}, respectively, between which the critical curve separating the two areas is located.
        There is an asymptote (orange-dotted line) $\mu^{-1}\approx 0.437$ for the critical curve when $\sigma \to +\infty$.
        The magenta circles represent the simulation results of $\mu_{\rm C}$ on a complete graph with $N=100$.
    }
    \label{fig:pareto_phase}
\end{figure}

Similar to \autoref{fig:gamma_sigma} for the gamma distribution, \autoref{fig:pareto_sigma} shows the mean-field solutions and simulation results of $\left<x\right>$ vs. $\mu^{-1}$ for $\sigma=1$ under Pareto distribution.
The mean-field solutions show a good agreement with the simulation results on, with a critical point $\mu_{\rm C}^{-1}\approx 0.524$ and $\mu_{\rm C}^{-1}\approx 0.514$, respectively.

\autoref{fig:pareto_phase} shows the phase diagram with respect to $(\sigma, \mu^{-1})$ under the Pareto distribution~\eqref{eq:pareto} (see also Sec.~V of SM for a comparison between the mean-field solutions and simulation results).
The upper bound line is
\begin{align}
    \frac{3125(\mu^2+\sigma^2)^2}{768\mu^2}=\mu^4-\sigma^2\mu^2-2\sigma^4-2\sigma^3\sqrt{\mu^2+\sigma^2},
\end{align}
calculated from Eq.~\eqref{eq:bounds} and $\gamma^3=\alpha\beta_{\rm min}^3/(\alpha-3)$.
As distinct from the gamma distribution case, the critical value remains finite even when $\sigma \to +\infty$, leading to a qualitatively different phase diagram.
As $\sigma\to +\infty$ yields $\alpha=2$ and a finite $\rho(\beta)$, an asymptote appears (orange-dotted line in \autoref{fig:pareto_phase}), which consists with Eq.~\eqref{eq:asymptotic_mu}.
With a sufficient number of low temperature links reflecting the heavy-tail structure, the polarized state is stabilized even with increasing $\sigma$ as predicted by the mean-field theory.

\subsection{Double-delta distribution}
\label{subsec_double_delta}
\begin{figure}
    \includegraphics[width=0.95\columnwidth]{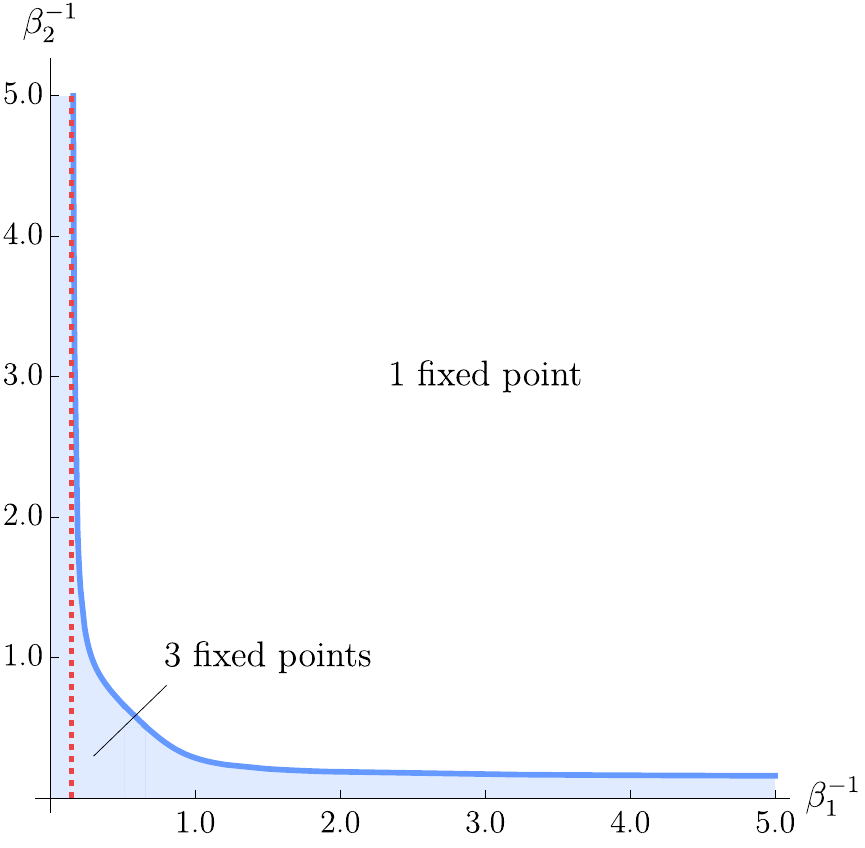}
    \centering
    \caption{
        Phase diagram with respect to $(\beta_1^{-1}, \beta_2^{-1})$ for the double-delta distribution with $q=0.5$.
        The diagram is separated by three fixed-points area and only one trivial fixed-point ($\left<x\right>=0$) area.
        The red-dotted line represents the critical value of $\beta_1\approx 6.86598$, above which there is no phase transition regardless of the value of $\beta_2$.
    }
    \label{fig:double_delta}
\end{figure}

Double-delta distribution can represent the case with heterogeneous but discrete social temperatures in a complete graph, where $T_{ij}$ takes two different values $T_1$ and $T_2$ with ratio $q$ and $1-q$, respectively.
In this case, the distribution of $\beta$ is given by
\begin{equation}
    \rho(\beta) = q \delta\left(\beta - \beta_1\right) + (1-q) \delta\left(\beta - \beta_2\right),
\end{equation}
where $\beta_1 \equiv (N-2)/T_1$ and $\beta_2 \equiv (N-2)/T_2$.
The self-consistent equation~\eqref{eq:self_consistent_supp} and the critical point~\eqref{eq:critical_point_supp} reduce to
\begin{align}
    F(x) &= q \tanh\left(\beta_1  x^2\right) + (1-q) \tanh\left(\beta_2 x^2\right)-x\notag\\
    &=0,\\
    G(x) &= 2 q \beta_1 x  \sech^2\left(\beta_1 x^2\right)+ 2 (1-q) \beta_2 x  \sech^2\left(\beta_2 x^2\right)-1\notag\\
    &=0,
\end{align}
respectively.

We can find
\begin{align}
    \frac{d\beta_1}{d\beta_2}&=\frac{\frac{\partial F}{\partial x}\frac{\partial G}{\partial \beta_2}-\frac{\partial F}{\partial \beta_2}\frac{\partial G}{\partial x}}{\frac{\partial F}{\partial \beta_1}\frac{\partial G}{\partial x}-\frac{\partial F}{\partial x}\frac{\partial G}{\partial \beta_1}}\notag\\
    &=-\frac{\frac{\partial F}{\partial \beta_2}}{\frac{\partial F}{\partial \beta_1}}\notag=-\frac{(1-q)\sech^2\left(\beta_2 x^2\right)}{q\sech^2\left(\beta_1 x^2\right)}\leq 0,
\end{align}
for the critical line in the phase diagram (\autoref{fig:double_delta}).
This implies a critical value of $\beta_1$, above which there is no phase transition regardless of the value of $\beta_2$.
The critical value can be obtained by solving the equations that putting $\beta_2=0$:
\begin{align}
    \left< x \right> &= q \tanh\left(\beta_{1} \left< x \right>^2\right),\\
    1 &= 2 q \beta_{1} \left< x \right> \sech^2\left(\beta_{1} \left< x \right>^2\right).
\end{align}
The similar critical value for $\beta_2$ can also be obtained by exchanging the subscripts ``$1$'' to ``$2$'', and replacing ``$q$'' with ``$(1-q)$'' above.

The critical value, which depends on the ratio $q$, implies that the polarized state can be stabilized if there is enough ratio of links, of which the temperature is low enough.
This result highlights a key mechanism underlying the effect of heterogeneity: even a small fraction of low temperature links can dominate the collective behavior and stabilize the polarized phase.
This provides an intuitive explanation for the difference between light-tailed and heavy-tailed distributions observed in the phase diagrams.

Figure~\ref{fig:double_delta} shows the phase diagram with respect to $(\beta_1^{-1}, \beta_2^{-1})$ for $q=0.5$.
As shown in the figure, we can find the critical curve
that separates the phase with three fixed points and the phase with only one trivial fixed point $\left<x\right>=0$.
The red-dotted line marks the critical value $\beta_1\approx 6.86598$, beyond which no phase transition occurs for any value of $\beta_2$.

Such a critical value also exists for decreased multi-delta distributions $\displaystyle\rho(\beta) = \sum_{i}q_i \delta(\beta-\beta_i)$ with more than two peaks, which can be obtained similarly by ignoring the contributions from other peaks.

\section{Concluding Remarks}
\label{sec_con}
In this work, we developed a generalized framework for Heider balance dynamics in which each interpersonal relation is assigned its own social temperature.
By formulating the model through heterogeneous social temperatures, we established a mean-field description that connects the classical theory of structural balance with concepts from disordered and higher-order interacting systems.

This perspective revealed how collective opinion states depend not only on the average level of social volatility but on the entire distribution of temperature heterogeneity.
In particular, the tail structure of the inverse temperature distribution leads to qualitatively distinct phase behaviors.
While increasing dispersion generally destabilizes polarized states for light-tailed distributions, heavy-tailed distributions admit a robust regime in which polarized states persist whenever the mean temperature is low enough, regardless of the level of fluctuations.
This difference between the behaviors in the phase diagram reveals the role of low temperature links in Heider dynamics, which stabilize the polarized phase if their ratio is sufficient.
In fact, even a simple two temperature, double-delta distribution model can stabilize the polarized state if one temperature is low enough, as the result in Sec.~\ref{subsec_double_delta}.

Moreover, we also derived the universal bounds for the critical transition, where the homogeneous-temperature limit, corresponding to a delta distribution, provides a universal lower bound for the critical mean of general inverse temperature distributions.

The present results open up several directions for future research.
A natural next step is to move beyond the mean-field approximation and investigate the role of fluctuations.
Analytical approaches from statistical field theory or numerical techniques could shed light on detailed phase structures in this setting~\cite{bagherikalhor2021heider,umantsev2012field,babaee2022individual}.

Another promising direction is the introduction of external fields acting on links or nodes~\cite{burda2025heiderbalancesquarelattice,PhysRevLett.133.127402,burda2025relationshipheiderlinksising}.
Allowing these fields to vary across the network would make it possible to model polarized or asymmetric environments and to study how structural balance is reshaped when individuals are systematically influenced by heterogeneous external factors.

Finally, applying the present framework to networks with nontrivial topology would provide further insight into how structural balance unfolds in realistic social systems~\cite{belaza2019social}.
On such graphs, the distribution of triadic motifs and local neighborhood structures differs significantly from that of complete graphs, and these differences may interact in meaningful ways with heterogeneous temperature distributions.

\begin{acknowledgements}
This work was supported by JST SPRING, Grant Number JPMJSP2108.
\end{acknowledgements}

\appendix
\section{Derivation of the self-consistent equation}
\label{app:self}
Here, we derive the self-consistent equation~\eqref{eq:self_consistent} in Sec.~\ref{subsec_self}.
We can approximate the effective field $h_{\rm eff}\left(x_{ij}\right)$ acting on $x_{ij}$ as
\begin{equation}
    h_{\rm eff}\left(x_{ij}\right) \equiv -\left< x \right>^2 M_{ij} x_{ij},
\end{equation}
from the local Hamiltonian~\eqref{eq:local_hamiltonian} by ignoring fluctuations and correlations.
Here, $M_{ij}=N-2$ is the number of the common neighbors between nodes $i$ and $j$, as we mainly consider a complete graph with $N$ nodes.

The partition function $Z_{ij}$ for the opinion variable $x_{ij}$ under the effective field is given by
\begin{align}
    Z_{ij} &= \sum_{x_{ij}=\pm 1} \exp\left[-T_{ij}^{-1} h_{\rm eff}\left(x_{ij}\right)\right]\notag\\
    &=\sum_{x_{ij}=\pm 1}\exp\left[\frac{M_{ij}}{T_{ij}}\left< x \right>^2x_{ij}\right]\notag\\
    &= 2\cosh\left(\frac{M_{ij}}{T_{ij}} \left< x \right>^2\right),
\end{align}
with respect to its own social temperature $T_{ij}$.
We can define
\begin{equation}
    \beta_{ij} \equiv \frac{M_{ij}}{T_{ij}}.
\end{equation}
with which the structure of the graph ($M_{ij}$) and the social temperatures ($T_{ij}$) are captured together as heterogeneous positive definite ``coupling strengths'' $\beta_{ij}$.
This allows us to treat the system as if it has homogeneous fixed social temperatures ($T_{ij}=T$) but heterogeneous coupling strengths, which we may find in spin-glass systems~\cite{mezard1987spin,nishimori2001statistical}.
We can rewrite the partition function $Z_{ij}$ as
\begin{equation}
    Z_{ij} = 2\cosh\left(\beta_{ij} \left< x \right>^2\right).
\end{equation}
The mean value $\bar{x}_{ij}$ of the opinion variable under the effective field is given by
\begin{align}
    \bar{x}_{ij} = \frac{\partial Z_{ij}}{\partial \left(\beta_{ij}\left<x\right>^2\right)}= \tanh\left(\beta_{ij} \left< x \right>^2\right).
\end{align}
Taking the average over all links in the network, we obtain the self-consistent equation for the average opinion $\left< x \right>$:
\begin{equation}
    \left< x \right> = \left< \tanh\left(\beta_{ij} \left< x \right>^2\right) \right>.
\end{equation}
If $\beta_{ij}$ is distributed as $\rho(\beta)$, the self-consistent equation can be expressed as
\begin{equation}
    \left< x \right> = \int_0^{+\infty} \tanh\left(\beta \left< x \right>^2\right)\rho(\beta)d\beta,
\end{equation}
which is Eq.~\eqref{eq:self_consistent}.

\section{Derivation of the bounds for the critical mean}
\label{app_bounds}
In this part, we derive the universal bounds~\eqref{eq:bounds} for the critical mean $\mu_{\rm C}$.
We rewrite the self-consistent equation~\eqref{eq:self_consistent} and the critical point equation~\eqref{eq:critical_point} as the root finding problems of the functions $F(x)$ and $G(x)$, respectively:
\begin{align}
    F(x)&\equiv \int_0^{+\infty} \tanh\left(\beta x^2\right)\rho(\beta)d\beta - x = 0,\label{eq:self_consistent_supp}\\
    G(x)&\equiv \int_0^{+\infty} 2 \beta x \sech^2\left(\beta x^2\right)\rho(\beta)d\beta - 1 = 0.\label{eq:critical_point_supp}
\end{align}

We first derive the lower bound.
Applying Jensen's inequality~\cite{dragomir1994some} to Eq.~\eqref{eq:self_consistent_supp} leads to
\begin{align}
    x=\int_0^{+\infty} \tanh\left(\beta x^2\right)\rho(\beta)d\beta \leq \tanh\left(\mu x^2\right),
\end{align}
which implies a lower bound
\begin{align}
    \mu \geq \frac{\arctanh x}{x^2}.\label{eq:lower_mu_supp}
\end{align}
The equality holds if and only if $\beta$ is constant almost everywhere, leading to a delta distribution for $\rho(\beta)$.
Thus, we can obtain the lower bound $C_*$ for $\mu_{\rm C}$ by solving the critical point for the delta distribution $\rho(\beta)=\delta(\beta-C_*)$:
\begin{align}
    \tanh\left(C_* x_*^2\right)-x_*&=0,\\
    2C_*x_*\sech^2\left(C_* x_*^2\right)-1&=0,
\end{align}
resulting $C_*\approx 1.716$ and $x_* \approx 0.796$.
The $C_*$ and $x_*$ also align with Eq.~\eqref{eq:lower_mu_supp}, as the function $x \mapsto x^{-2}\arctanh x$ ($x \in [0,1]$) takes the minimum value $C_*$ at $x_*$

Before deriving the upper bound controlled by the third moment $\gamma^3$ of the distribution $\rho(\beta)$, we first provide the range of possible values for it.
By applying Jensen's inequality~\cite{dragomir1994some} to the convex function $\beta \mapsto \beta^{3}$, we have
\begin{equation}
    \gamma^3 = \int_0^{+\infty} \beta^3 \rho(\beta)d\beta \geq \left(\int_0^{+\infty} \beta \rho(\beta)d\beta\right)^3 = \mu^3,
\end{equation}
which leads to $\gamma \geq \mu$.

We apply the following inequalities to derive the upper bound:
\begin{align}
    \tanh\left(u\right) &> u - \frac{1}{3} u^3,\\
    \sech^2\left(u\right) &> 1 - u^2,
\end{align}
for $u>0$.
Substituting them into Eqs.~\eqref{eq:self_consistent_supp} and~\eqref{eq:critical_point_supp}, we have
\begin{align}
    F(x) &> \int_0^{+\infty} \left(\beta x^2 - \frac{1}{3} \beta^3 x^6\right)\rho(\beta)d\beta - x\notag\\
    &= \mu x^2 - \frac{1}{3} \gamma^3 x^6 - x,\label{eq:F_lower_bound}\\
    G(x) &> \int_0^{+\infty} 2 \beta x \left(1 - \beta^2 x^4\right)\rho(\beta)d\beta - 1\notag\\
    &= 2 \mu x - 2 \gamma^3 x^5 - 1.\label{eq:G_lower_bound}
\end{align}
To ensure the existence of the critical point for Eqs.~\eqref{eq:self_consistent_supp} and~\eqref{eq:critical_point_supp}, it is sufficient to require the right-hand sides of Eqs.~\eqref{eq:F_lower_bound} and~\eqref{eq:G_lower_bound} to have roots.
By solving the equations
\begin{align}
    \mu x^2 - \frac{1}{3} \gamma^3 x^6 - x &= 0,\\
    2 \mu x - 2 \gamma^3 x^5 - 1 &= 0,
\end{align}
we obtain a line
\begin{equation}
    \tilde{l}(\mu, \gamma) \equiv \left(\frac{5}{4}\right)^5\gamma^3-\frac{3}{4}\mu^5=0,\label{eq:lower_curve}
\end{equation}
and a corresponding point $\tilde{x}^{\rm C}$.
Due to the point $\tilde{x}^{\rm C}$ being
\begin{align}
    F(\tilde{x}^{\rm C})>0,\notag\\
    G(\tilde{x}^{\rm C})>0,
\end{align}
the line $\tilde{l}(\mu, \gamma)=0$ must fully lie in the phase with three fixed points, giving the upper bound for $\mu_{\rm C}$:
\begin{equation}
    \mu_{\rm C} < \frac{5}{4}\left(\frac{4}{3}\gamma^3\right)^{\frac{1}{5}}.
\end{equation}

Finally, we obtain the bounds Eq.~\eqref{eq:bounds} in Sec.~\ref{subsec_bounds}:
\begin{equation}
     C_* \leq \mu_{\rm C} < \frac{5}{4}\left(\frac{4}{3}\gamma^3\right)^{\frac{1}{5}},
\end{equation}
for $\gamma \geq \displaystyle\frac{25}{16}\sqrt{\frac{5}{3}}$.
The range of $\gamma$ is required due to $\gamma\geq\mu$ derived above, which comes from $\displaystyle \gamma \geq \frac{5}{4}\left(\frac{4}{3}\gamma^3\right)^{\frac{1}{5}}$.

\section{Proof of the asymptotic behavior for the critical mean}
\label{app_asym}
The proof of Eq.~\eqref{eq:asymptotic_mu} can be considered in the following way.
We first consider the case where the distribution $\rho(\beta)$ vanishes as $\lambda \to +\infty$ for all $\beta>0$, which means that for a distribution family $\rho_\lambda(\beta)$ parameterized by $\lambda$,
the integration on a finite interval $[0,B]$ with respect to $\beta$ is $0$ in the limit $\lambda \to +\infty$:
\begin{equation}
    \forall B>0, \quad \int_0^{B} \lim_{\lambda \to +\infty} \rho_\lambda(\beta)d\beta = \int_0^{B} \rho_*(\beta)d\beta =0.\label{eq:vanish}
\end{equation}

We can show that $\displaystyle \lim_{\lambda \to +\infty}x^{\rm C}_{\lambda} \to 0$ for the corresponding critical point.
Assume that there exists a finite $x_*>0$ such that $\displaystyle \lim_{\lambda \to +\infty}x^{\rm C}_{\lambda} = x_*$.
With this finite $x_*$, we can find a large enough $B$ such that $\tanh\left(\beta x_{*}^2\right)$ is arbitrarily close to $1$ if $\beta > B$:
\begin{align}
    \forall \delta > 0, \quad \exists B > 0, \quad &\text{s.t.} \quad \forall \beta > B, \notag\\
    \tanh\left(\beta x_{*}^2\right) &> 1 - \delta,
\end{align}
while for the same $B$, derived from Eq.~\eqref{eq:vanish}, there exists a sufficiently small $\epsilon$ such that
\begin{equation}
    \int_B^{+\infty} \rho_*(\beta)d\beta > 1-\epsilon.
\end{equation}
Thus, we have
\begin{align}
    F(x_*) &= \int_0^{+\infty} \tanh\left(\beta x_*^2\right)\rho_*(\beta)d\beta - x_*\notag\\
    &> \int_B^{+\infty} \tanh\left(\beta x_*^2\right)\rho_*(\beta)d\beta - x_*\notag\\
    &> (1-\delta)(1-\epsilon) - x_*,
\end{align}
when substituting $x_*$ and $\rho_*(\beta)$ into Eq.~\eqref{eq:self_consistent_supp}.
This enforces
\begin{equation} 
    x_*=1,
\end{equation}
as $\delta, \epsilon \to 0^+$.
However, substituting $x_*=1$ and $\rho_*(\beta)$ into Eq.~\eqref{eq:critical_point_supp} leads to
\begin{align}
    G(x_*)=G(1) = \int_0^{+\infty} 2 \beta  \sech^2\left(\beta\right)\rho_*(\beta)d\beta - 1<0,
\end{align}
which contradicts the definition of the critical point.
Here, we applied the fact that the upper bound of the function $\beta \mapsto \beta \sech^2(\beta)$ is around $0.448$ at $\beta \approx 0.772$.

Thus, we need to assume $\displaystyle \lim_{\lambda \to +\infty}x_\lambda^{\rm C}=0$.
By replacing the inequality into equality of Eqs.~\eqref{eq:F_lower_bound} and~\eqref{eq:G_lower_bound}, which here can be regarded as expansions with respect to $x_\lambda^{\rm C} \to 0$, we again find the same relationships as Eq.~\eqref{eq:lower_curve} for $x_\lambda^{\rm C}$, $\gamma_{\rm C}$ and $\mu_{\rm C}$.
We then derive $\left(x_\lambda^{\rm C}\right)^5\gamma_{\rm C}^{3}\sim O(1)$, which leads to
\begin{eqnarray}
    x_\lambda^{\rm C}\sim \gamma^{-\frac{3}{5}}_{\rm C}.
\end{eqnarray}
This relationship consists with the assumption $x_\lambda^{\rm C} \to 0$, as $\lambda \to +\infty$ in the vanishing limit of the distribution.
Moreover, we have $x_\lambda^{\rm C}\mu_{\rm C}\sim O(1)$, resulting in
\begin{eqnarray}
    x_\lambda^{\rm C}\sim \mu_{\rm C}^{-1},
\end{eqnarray}
which implies the divergence of $\mu_{\rm C}$:
\begin{equation}
    \lim_{\lambda\to +\infty}\mu_{\rm C} = +\infty.
\end{equation}

On the other hand, we can obtain a finite $\mu_{\rm C}$ from Eqs.~\eqref{eq:self_consistent_supp} and~\eqref{eq:critical_point_supp} if $\rho(\beta)$ remains finite in the limit $\lambda \to +\infty$.
Finally, we derive the asymptotic behavior of $\mu_{\rm C}$ as Eq.~\eqref{eq:asymptotic_mu} in Sec.~\ref{subsec_asym}.

\bibliography{main}

\end{document}

%% file: fig.tex
\begin{tikzpicture}

\definecolor{cA}{rgb}{0.00,0.20,0.80}
\definecolor{cB}{rgb}{0.00,0.65,0.85}
\definecolor{cC}{rgb}{0.00,0.60,0.20}
\definecolor{cD}{rgb}{0.95,0.55,0.00}
\definecolor{cE}{rgb}{0.85,0.00,0.10}

\begin{scope}[local bounding box=graph]

\coordinate (P1) at (0.8, 1.2);
\coordinate (P2) at (3.2, 0.6);
\coordinate (P3) at (2.4, 2.8);
\coordinate (P4) at (0.6, 3.0);
\coordinate (P5) at (4.0, 2.4);

\draw[line width=0.95pt, draw=cB] (P1)--(P2);
\draw[line width=0.95pt, draw=cE] (P1)--(P3);
\draw[line width=0.95pt, draw=cA] (P1)--(P4);
\draw[line width=0.95pt, draw=cD] (P1)--(P5);

\draw[line width=0.95pt, draw=cC] (P2)--(P3);
\draw[line width=0.95pt, draw=cD] (P2)--(P4);
\draw[line width=0.95pt, draw=cA] (P2)--(P5);

\draw[line width=0.95pt, draw=cB] (P3)--(P4);
\draw[line width=0.95pt, draw=cC] (P3)--(P5);

\draw[line width=0.95pt, draw=cE] (P4)--(P5);

\foreach \P in {P1,P2,P3,P4,P5}{
  \fill (\P) circle (2.2pt);
}

\end{scope}

\coordinate (cbpos) at ($ (graph.east) + (0.5cm,0) $);

\node[anchor=west] at (cbpos) {%
\begin{tikzpicture}
\begin{axis}[
  width=0pt,
  height=3.6cm,
  hide axis,
  scale only axis,
  colorbar,
  point meta min=0,
  point meta max=1,
  colormap={hiContrast}{
    color=(cA)
    color=(cB)
    color=(cC)
    color=(cD)
    color=(cE)
  },
  colorbar style={
    width=0.38cm,
    ytick=\empty,
    ylabel={Social temperature},
    ylabel style={font=\small},
  },
]
\addplot[draw=none] coordinates {(0,0)};
\end{axis}
\end{tikzpicture}
};

\end{tikzpicture}

%% file: figb.tex
\begin{tikzpicture}

\definecolor{coldlink}{RGB}{52,120,246}   
\definecolor{hotlink}{RGB}{220,70,70}     
\definecolor{midlink}{RGB}{60,170,90}     

\coordinate (k1) at (0,1.8);
\coordinate (i1) at (-1.1,0);
\coordinate (j1) at (1.1,0);

\node[below left] at (i1) {$i$};
\node[below right] at (j1) {$j$};
\node[above] at (k1) {$k$};

\draw[line width=0.95pt,coldlink] (i1)--(k1);
\draw[line width=0.95pt,midlink]  (j1)--(k1);

\draw[line width=0.95pt,hotlink] (j1)--(i1);

\node at (-0.75,0.95) {$+$};
\node at (0.75,0.95) {$+$};
\node at (0,-0.22) {$x_{ij}$};

\draw[->,thick] (1.5,1.1) -- (3.5,2.1)
    node[midway,above=3pt] {$p_{ij}$};

\draw[->,thick] (1.5,0.7) -- (3.5,-0.3)
    node[midway,below=3pt] {$1-p_{ij}$};

\coordinate (k2) at (4.7,3.5);
\coordinate (i2) at (3.6,1.7);
\coordinate (j2) at (5.8,1.7);

\node[below left] at (i2) {$i$};
\node[below right] at (j2) {$j$};
\node[above] at (k2) {$k$};

\draw[line width=0.95pt,coldlink] (i2)--(k2);
\draw[line width=0.95pt,midlink]  (j2)--(k2);
\draw[line width=0.95pt,hotlink]  (j2)--(i2);

\node at (3.97,2.63) {$+$};
\node  at (5.38,2.63) {$+$};
\node  at (4.7,1.47) {$x_{ij}=1$};

\coordinate (k3) at (4.7,0.7);
\coordinate (i3) at (3.6,-1.1);
\coordinate (j3) at (5.8,-1.1);

\node[below left] at (i3) {$i$};
\node[below right] at (j3) {$j$};
\node[above] at (k3) {$k$};

\draw[line width=0.95pt,coldlink] (i3)--(k3);
\draw[line width=0.95pt,midlink]  (j3)--(k3);
\draw[line width=0.95pt,hotlink] (j3)--(i3);

\node at (3.97,-0.17) {$+$};
\node  at (5.38,-0.17) {$+$};
\node  at (4.7,-1.33) {$x_{ij}=-1$};

\end{tikzpicture}